\documentclass[12pt]{article}

\usepackage{graphicx}

\newcommand{\tr}{\mbox{Tr} \, }
\newcommand{\ket}[1]{\left | #1 \right \rangle}
\newcommand{\bra}[1]{\left \langle #1 \right |}
\newcommand{\amp}[2]{\left \langle #1 \left | #2 \right. \right \rangle}

\newcommand{\proj}[1]{\ket{#1} \! \bra{#1}}

\newcommand{\ave}[1]{\left \langle #1 \right \rangle}
\newcommand{\superop}{{\cal E}}
\newcommand{\truncate}{{\cal T}}

\newcommand{\hilbert}{{\cal H}}
\newcommand{\relent}[2]{{\cal D}\left ( #1 || #2 \right )}

\newcommand{\zef}{\mbox{\sf zef} }
\newcommand{\subzef}{\mbox{\scriptsize \sf zef} }
\newcommand{\lenobs}{\mbox{$\Lambda$} }
\newcommand{\biglenobs}{\mbox{\boldmath $\Lambda$}}
\newcommand{\prob}[1]{\mbox{Pr} \left (  #1  \right )}
\newcommand{\opnorm}[1]{\mbox{$\left \| #1 \right \| $}}

\newcommand{\zshortish}{_{\scriptscriptstyle (\preceq \ell)}}

\newcommand{\zlong}{_{\scriptscriptstyle (\succ \ell)}}
\newcommand{\oshort}{_{\scriptscriptstyle (< \ell)}}

\newcommand{\olongish}{_{\scriptscriptstyle (\geq \ell)}}

\begin{document}

\title{Indeterminate-length quantum coding}

\author{Benjamin Schumacher$^{(1)}$
        and Michael D. Westmoreland$^{(2)}$}
\maketitle
\begin{center}
{\sl
$^{(1)}$Department of Physics, Kenyon College, Gambier, OH 43022 USA \\
$^{(2)}$Department of Mathematical Sciences, Denison University,
 Granville, OH  43023 USA }
\end{center}

\section*{Abstract}

The quantum analogues of classical variable-length codes are {\em
indeterminate-length} quantum codes, in which codewords may exist
in superpositions of different lengths.  This paper explores some
of their properties.  The length observable for such codes is
governed by a quantum version of the Kraft-McMillan inequality.
Indeterminate-length quantum codes also provide an alternate
approach to quantum data compression.

\section*{Introduction}

The development of quantum information theory is a striking example
of the fruitful hybridization of two well-established disciplines.
Both quantum mechanics and information theory have a rich set of
concepts and a powerful toolbox of mathematical techniques.  Their
combination is yielding powerful insights into the physical meaning
of ``information'' \cite{chbinphystoday,mikeandike}.

One approach to this exploration is to begin with an idea of
``classical'' information theory and investigate how this idea
must be re-interpreted or modified to fit into the quantum
information framework.  Ideas of fidelity,
quantum data compression \cite{qcompress},
quantum error correcting codes \cite{qerror},
and the capacities of various quantum channels \cite{qcapacity}
can all be viewed in this light.

A basic idea in the classical theory of data compression is the
idea of a variable-length code.  A variable-length code assigns
to different messages codewords consisting of different numbers
of symbols.  If shorter codewords are used for more common messages
and longer ones for less common messages, the average codeword
length can be made shorter than would be possible using a
fixed-length code.  (Natural languages take advantage of this
idea.  Common words like ``the'' are often very short, while
unusual words like ``sesquipedalian'' are longer.)

However, the original development of quantum data compression
followed a different route, parallel to the classical development
based on ``typical sequences''.  This left open the question
of whether there was a quantum analogue to classical
variable-length coding.  Because a quantum code must allow
superpositions of different codewords---including superpositions
of codewords of {\em different lengths}---the quantum version
would best be termed an {\em indeterminate}-length quantum code.

One of us \cite{beninsantafe} made a preliminary investigation
of this idea several years ago.  Subsequently, Braunstein et al.
\cite{qhuffman} presented a quantum analogue to classical Huffman
coding.  Because a general understanding of indeterminate-length
quantum codes was not available then, Braunstein et al. were
led to construct their code in an unnecessarily inefficient way.
(See the discussion in Section~\ref{discuss-sec}, below.)  More
recently, Chuang and Modha have developed a quantum version of
arithmetic coding as a route to quantum data compression
\cite{chuang}.  Bostr\"{o}m has also investigated indeterminate-length
codes in connection with lossless quantum coding \cite{bostrom}.

Our aim in this paper is to outline a general theory of
indeterminate-length quantum codes, including their
application to quantum data compression.

We will first sketch a framework for discussing such codes.  Each code
will have a ``codeword length'' observable $\lenobs$ with
integer eigenvalues; allowable codewords include not only
length eigenstates but arbitrary superpositions of them.
The key requirement is that such codes be ``condensable''---that is,
that the individual codewords can be assembled into a string
by means of a unitary operation.  This condition leads us
to prove a quantum version of the Kraft-McMillan inequality.
Among the condensable codes are those that satisfy a quantum
``prefix-free'' condition, and we show (by giving an explicit
condensation algorithm) that all such
codes are condensable.  We also show how classical
variable-length codes can be used to construct quantum
indeterminate-length codes with analogous properties.

We next turn to the use of indeterminate-length codes for
quantum data compression.  We achieve quantum data compression
by taking a condensed string of $N$ codewords (in general no
shorter than $N$ times the largest eigenvalue of $\lenobs$)
and truncating it after the first $N \ell$ qubits, thus
using only $\ell$ qubits per input codeword.  We show
that the average $\ave{l}$ of the codeword length observable
$\lenobs$ is the necessary and sufficient value of $\ell$
to achieve high fidelity for this process.  It turns out that
$\ave{l}$ is related to the quantum entropy $S$ of the
quantum information source, and from this relation we are able
to arrive at the noiseless quantum coding theorem.

\section{Indeterminate-length codes}

\subsection{Zero-extended forms}

In a quantum code, codewords are states of finite strings of qubits.
Superpositions of codewords are also valid codewords, and to maintain
high fidelity we must preserve the coherence of these superpositions
in our coding and decoding processes.

We wish to create a code in which different codewords have different
{\em lengths}---that is, they involve different numbers of qubits.
But how do we make sense of this idea?  We'll begin by considering
{\em zero-extended forms} (\zef) of the codewords.  For \zef codewords,
we imagine that the codewords are sitting at the beginning of a
qubit register of fixed length, with $\ket{0}$'s following.  These
codewords span a subspace of the Hilbert space of register states.

Our first essential requirement is that the codewords carry their
own length information.  That is, we require that there is a
``length'' observable \lenobs on the \zef codeword subspace with the
following two properties:
\begin{itemize}
	\item  The eigenvalues of \lenobs are $1, \ldots , l_{max}$,
		where $l_{max}$ is the length of the register.
	\item  If $\ket{\psi_{\subzef}}$ is an eigenstate of \lenobs
		with eigenvalue $l$, then it has the form
		\begin{equation}
		\ket{\psi_{\subzef}} = \ket{\psi^{1 \cdots l}
					0^{l+1 \cdots l_{max}}} .
		\end{equation}
		In other words, the last $l_{max} - l$ qubits
		in the register are in the state $\ket{0}$ for a
		\zef codeword of length $l$.
\end{itemize}
The length observable $\lenobs$ was also considered in \cite{bostrom}.

For each $l = 1, \ldots , l_{max}$, we let $d_{l}$ be the dimension
of the subspace spanned by the \lenobs-eigenstates with eigenvalue $l$.
Denote the projection onto this subspace by $\pi_{l}$.  Then
$\tr \pi_{l} = d_{l}$.

\subsection{Condensable codes}

We want to be able to make use of the comparative shortness
of some codewords by ``packing'' the codewords together, eliminating
the trailing zeroes that ``pad'' the ends of the \zef codewords.
But this must be a process that maintains quantum coherences
in superpositions of codeword states---that is, it must be
described by a {\em unitary} transformation.  Furthermore, we wish
to be able to coherently pack together any number of codewords.

We say that a code is {\em condensable} if the following
condition holds:  For any $N$, there is a unitary
operator $U$ (depending on $N$) that maps
\begin{equation}
	\underbrace{\ket{\psi_{1,\subzef}} \otimes \cdots \otimes
			\ket{\psi_{N,\subzef}}}_{N l_{max} \mbox{qubits}}
		\rightarrow
	\underbrace{\ket{\Psi_{1 \cdots N,\subzef}}}_{N l_{max} \mbox{qubits}}
\end{equation}
with the property that, if the individual codewords are all length
eigenstates, then $U$ maps the codewords to a \zef string of the
$N l_{max}$ qubits---that is, one with $\ket{0}$'s after the first
$L = l_{1} + \cdots + l_{N}$ qubits:
\begin{equation}
	\ket{\psi_{1}^{1 \cdots l_1} 0^{l_1 + 1 \cdots l_{max}}}
		\otimes \cdots \otimes
	\ket{\psi_{N}^{1 \cdots l_{N}} 0^{l_N + 1 \cdots l_{max}}}
	\rightarrow
	\ket{\Psi^{1 \cdots L}
			0^{L + 1 \cdots Nl_{max}}} .  \label{condense-eqn}
\end{equation}
This process is called {\em condensation}.  Since every codeword is
a superposition of length eigenstates, it suffices to specify how the
condensation process functions for such codewords.

	Note that we have made no assumptions about the details of the
condensation process.  In the most straightforward case, condensation
would be accomplished by concatenation of the codewords.  The
condensed state in Equation~\ref{condense-eqn} would be of the form
\begin{equation}
\ket{\Psi^{1 \cdots L} 0^{L + 1 \cdots Nl_{max}}} =
	\ket{\psi_{1}^{1 \cdots l_1} \, \cdots \, \psi_{N}^{L-l_N+1 \cdots L}
			\, 0^{L + 1 \cdots Nl_{max}}} .
\end{equation}
This special type of condensation is called {\em simple condensation},
and those codes whose codewords can be condensed in this way are said
to be {\em simply condensable} codes.  Obviously, all simply condensable
codes are condensable; but the converse is not true.

The condensability condition is phrased as an ``encoding''
requirement, but the unitary character of the packing
process automatically yields a decoding condition---we can
unpack a condensed string by applying the $U^{-1}$ transformation.

It is interesting to compare the analogous classical situation.
Classical codewords in a variable-length code can always be
concatenated into a ``packed'' string.  Only for uniquely
decipherable codes is this packing reversible.  In the quantum
case, since arbitrary superpositions of codewords are also legal
codewords, the concatenation process itself must be unitary.
This automatically implies that it can be reversed.

\subsection{The quantum Kraft-McMillan inequality}

Given that the codewords carry their own length information and
form a condensable code, we next derive a condition on the codeword
length observable.  Fix a value of $N$ and consider
all codeword strings that have given values of $l_1, l_2, \ldots, l_N$.
These states lie in a subspace of dimension
$d_{l_1} d_{l_2} \cdots d_{l_N}$, and all of them are mapped by $U$
into something of the form $\ket{\Psi^{1 \cdots L}
\, 0^{L + 1 \cdots Nl_{max}}}$.

Next, imagine strings of codewords with different lengths
$l_1', l_2', \ldots, l_N'$, but whose lengths sum to the same
total length:  $L' = L$.  The space spanned by these
has dimension $d_{l_1'} d_{l_2'} \cdots d_{l_N'}$ and is orthogonal
to the previous space.  We can consider all such combinations of
lengths that sum to the same $L$.  Each of these states
maps under $U$ to something of the form
$\ket{\Psi^{1 \cdots L}
\, 0^{L + 1 \cdots Nl_{max}}}$, so we obtain
\begin{displaymath}
\left ( \mbox{
\parbox{1.7in}{\raggedright dimension of space containing
	all codeword strings with the same $L$}
} \right )    \leq    \left ( \mbox{
\parbox{1.7in}{\raggedright dimension of space containing all strings
$\ket{\Psi^{1 \cdots L}
\, 0^{L + 1 \cdots Nl_{max}}}$}
}\right )
\end{displaymath}
\begin{displaymath}
\sum_{l_1+\cdots+l_N=L} d_{l_1} \cdots d_{l_N}
  \leq   2^{L} .
\end{displaymath}
It follows that
\begin{displaymath}
2^{-L} \sum_{l_1+\cdots+l_N=L} d_{l_1} \cdots d_{l_N} =
\sum_{l_1+\cdots+l_N=L}
	\left ( 2^{-l_1} d_{l_1} \right ) \cdots
	\left ( 2^{-l_N} d_{l_N} \right ) \leq 1 .
\end{displaymath}
There are at most $N l_{max}$ possible values of $L$.  If we
sum both sides of this equation over those values, the resulting
sum on the left-hand side will include all possible values of
$l_1 , \ldots , l_N$.  Therefore,
\begin{displaymath}
\sum_{l_1,\ldots,l_{max}} \left ( 2^{-l_1} d_{l_1} \right ) \cdots
	\left ( 2^{-l_N} d_{l_N} \right ) =
	\left ( \sum_{l} 2^{-l} d_{l} \right )^N \leq N l_{max} .
\end{displaymath}
This is of the form $K^{N} \leq N l_{max}$.  If $K > 1$, then this
inequality must be violated for sufficiently large $N$.  Thus, we
conclude that $K \leq 1$.  But
\begin{displaymath}
	K =  \sum_{l} 2^{-l} d_{l}
	  =  \sum_{l} 2^{-l} \tr \pi_{l}
	  =  \tr \left ( \sum_{l} 2^{-l} \pi_{l} \right ) .
\end{displaymath}
This gives us our quantum version of the Kraft-McMillan inequality.
For any indeterminate-length quantum code that is condensable,
the length observable \lenobs on \zef codewords must satisfy
\begin{equation}
	\tr 2^{-\lenobs} \leq 1  \label{kraft}
\end{equation}
(where the trace is taken over the subspace of \zef codewords).

\subsection{Prefix-free codes}

An alternate condition that we might impose on our indeterminate-length
quantum code is that the code be {\em prefix-free}---informally, that
no initial segment of a \zef codeword is itself a codeword.
In the next section, we will show that all prefix-free codes
are simply condensable.
In this section, we will discuss the meaning of the prefix-free
condition and show that any condensable code can be transformed
into a prefix-free code with the same length characteristics.

Suppose $\ket{\psi_1}$ and $\ket{\psi_2}$ are length eigenstate
\zef codewords with lengths $l_1$ and $l_2$, respectively; and
further suppose that $l_2 > l_1$.  These states have the form
\begin{eqnarray}
\ket{\psi_1}  & = &
	\ket{\psi^{1 \cdots l_1}_1 0^{l_1+1 \cdots l_{max}}} \nonumber \\
\ket{\psi_2}  & = &
	\ket{\psi^{1 \cdots l_2}_2 0^{l_2+1 \cdots l_{max}}} .
\end{eqnarray}
For the codeword $\ket{\psi_1}$, the quantum state of the first
$l_1$ qubits of the register is just the pure state
$\ket{\psi^{1 \cdots l_1}_1}$.  For the codeword
$\ket{\psi_2}$, the first $l_1$ qubits may be in a
mixed state, described by the density operator
\begin{eqnarray}
\rho^{1 \cdots l_1}_2  & = &
    \tr_{l_1 + 1 \cdots l_{max}} \proj{\psi_2} \nonumber  \\
    & = &
    \tr_{l_1 + 1 \cdots l_2} \proj{\psi^{1 \cdots l_2}_2} .
\end{eqnarray}
We say that our code is {\em prefix-free} if, for all such pairs
of codewords,
\begin{equation}
	\bra{\psi^{1 \cdots l_1}_1} \rho^{1 \cdots l_1}_2
		\ket{\psi^{1 \cdots l_1}_1} = 0 .
\end{equation}
In other words, the first $l_1$ qubits of a codeword of length $l_1$
have a state that is orthogonal to the (possibly mixed) state of
the first $l_1$ qubits of a codeword of length $l_2 > l_1$.

Another way of expressing this condition is to say that a length
eigenstate \zef codeword of length $l$ can be distinguished
from a codeword of greater length by making a measurement only
on the first $l$ qubits.  Shorter codewords can be ``recognized''
from shorter segments.  This means that the operator $\pi_l$, the
projection onto the subspace of length eigenstate \zef codewords
of length $l$, has the form
\begin{equation}
	\pi_l = \pi^{1 \cdots l} \otimes 1^{l+1 \cdots l_{max}} .
\end{equation}

Of course, actually to measure the codeword length $\lenobs$
would be disastrous, because such a real measurement would
destroy the coherence of superpositions of length eigenstates
without possibility of restoration.  The condensation process
must therefore not include any measurement of length information.
On the other hand, the process may include interactions by which,
at some intermediate stage, a quantum computer has
become entangled with codeword length information---provided
that, by the end of the computation, this entanglement has
been eliminated.  In Section~\ref{coherent-sec} we discuss
this in more detail.

A particularly simple way of generating a prefix-free quantum
code is to use a classical prefix-free code as a basis for
the \zef codeword subspace.  For example, the classical
codewords 0, 10, 110 and 111 form a prefix-free set.  The
corresponding quantum code can be specified by giving an
orthogonal basis of length eigenstate \zef codewords,
as follows:
\begin{center}
\begin{tabular}{lc}
	state		&	length	\\
	$\ket{000}$	&	1	\\
	$\ket{100}$	&	2	\\
	$\ket{110}$	&	3	\\
	$\ket{111}$	&	3
\end{tabular} .
\end{center}
The length observable $\lenobs$ for this code is
\begin{equation}
	\lenobs = \proj{000} + 2 \proj{100}
			+ 3 \proj{110} + 3\proj{111} .
\end{equation}
Of course, any superposition of these is also a \zef codeword,
though not necessarily a codeword of definite $\lenobs$.  It is
easy to verify that the code defined in this way satisfies the
criterion for a quantum ``prefix-free'' code.
The procedure illustrated here may be extended in the
obvious way to create a quantum prefix-free code from any
classical prefix-free code.

Suppose we have a indeterminate-length quantum code that
satisfies the quantum Kraft-McMillan inequality.
Then the space of \zef codewords of this code is spanned
by a basis of eigenstates of this code's length observable
$\Lambda$.  Let $\ket{\psi_{\subzef ; l,i}}$ be the
$i$th such basis vector with length eigenvalue $l$,
and let $n_l$ be the number of basis vectors that
have length $l$.
(Thus for a given $l$, $i$ ranges from 1 to $n_l$.)
The quantum Kraft-McMillan inequality (Eq.~\ref{kraft})
implies that
\begin{equation}
	\sum_{l} n_l \, 2^{-l} \leq 1 .  \label{classkraft}
\end{equation}

Given values of $l$ and $n_l$ that satisfy Eq.~\ref{classkraft},
we can construct a classical prefix-free
code with $n_l$ distinct codewords of length $l$ bits.  (In this
case, Eq.~\ref{classkraft} is just the classical Kraft inequality.)
Denote by $C_{l,i}$ the $i$th codeword of length $l$ bits in this
prefix-free code.  We use this classical prefix-free code to
create a quantum prefix-free code by constructing a basis of
length eigenstates, whose elements are
$\ket{C_{l,i} \, 0^{l+1 \cdots l_{max}}}$.

Now consider the mapping
\begin{equation}
	\ket{\psi_{\subzef ; l,i}} \rightarrow
		\ket{C_{l,i} \, 0^{l+1 \cdots l_{max}}} .
	\label{ccode2qcode}
\end{equation}
This is a mapping from orthogonal basis vectors to orthogonal basis
vectors that can be extended linearly to a unitary mapping $V$ on the
entire Hilbert space.  $V$ takes the original codewords
to prefix-free codewords in a length-preserving way---that is, the
length observable $\lenobs'$ of the prefix-free code is given by
$\lenobs' = V \lenobs V^{\dagger}$.  In short, any quantum code
that satisfies Equation~\ref{kraft}
can be unitarily mapped to a prefix-free quantum code with identical
length characteristics.

Are all prefix-free quantum codes condensable?  As we shall see in
Section~\ref{pf-is-simpcond-sec}, they are; but in order to show this,
we will have to give an explicit algorithm for a quantum computer to
condense the codewords of a prefix-free quantum code.  This algorithm
must maintain the coherence of superpositions of codewords of different
lengths.  Before we describe our algorithm, we will first discuss
some key characteristics of coherent information processing.

\subsection{Coherence and reversibility}  \label{coherent-sec}

We adopt a high-level model of a quantum computer, which could in
principle be implemented by a quantum Turing machine or an array
of quantum gates.  Our quantum computer contains several registers
of qubits, which initially hold \zef codewords from a
prefix-free quantum code.  The computer also includes a central
processing unit that contains various counters and pointers,
each of which can take on integer values (or superpositions of
these).  A system clock keeps track of the number of machine
cycles that have passed since the beginning of the computation.
(This clock may be treated as an entirely classical system;
its function is simply to control the execution of our quantum
program.)  Finally, the computer contains
an output ``tape'' of qubits (initially all
in the state $\ket{0}$) on which the condensed string is to be
written.

Our job is to write the input code words onto the output tape
in a way that preserves the coherence of superpositions of
different codewords, including superpositions of codewords
of different lengths.  This means that the operation of the
computer must be unitary.  We can guarantee this unitarity
if we satisfy certain conditions:
\begin{enumerate}
\item  {\em Reversibility.}  In a classical code,
	all codewords have a determinate length.  We can choose an
	orthogonal basis of length eigenstates to be ``quasi-classical'' input
	states of our computer.  (These states need not be fully classical---for
	example, the qubits in these codeword states may be entangled with each
	other.  However, each codeword in our basis has a determinate length.)
	We require that distinct ``quasi-classical''
	inputs lead to distinct final states of the computer.  This is
	essentially a requirement that the computation be {\em reversible}
	on these quasi-classical inputs  \cite{bennett, toffoli} .
\item  {\em Coherent computation.}  The computation includes no
	measurement or process in which the environment becomes entangled
	with the computer.  As a special case of this, we require that
	the computation end after exactly the same number of steps for
	any input codeword.  If the computation took more
	steps for longer codewords, the halting time of the computation would
	constitute a measurement of codeword length, and would destroy the
	coherence.
\item  {\em Localization of coherence in the output.}
	For any quasi-classical input, at
	the end of the computation all input registers and internal
	variables in the central processor have been reset to fixed
	values that are independent of the input.
	Only the output tape retains any information about the input.  This will
	guarantee that a superposition of quasi-classical inputs will not lead
	to entanglement between the output tape and the rest of the computer;
	the coherence will be localized in the output tape.
\end{enumerate}
A similar set of conditions is outlined in \cite{chuang}, where it is
used to specify quantum algorithms for data compression and for
quantum arithmetic coding.

The reversibility requirement ensures that an orthogonal basis of initial
states maps to an orthogonal basis of final states.  If the computation is
coherent, this map extends by linearity to a unitary evolution for the
computer's quantum state.  The final requirement guarantees that the
quantum information initially in the input registers can be recovered
from the condensed output tape alone.
We will discuss each of these requirements in turn.

Consider how our quantum computer acts on quasi-classical
(length eigenstate) inputs.
If we were to map out its algorithm as a flowchart, the requirement
of reversibility would impose two sorts of requirements.  First,
each individual operation on the data must be reversible.  Second,
the branches and joins in the flowchart must be specified in a
reversible way.

A branch can be pictured in this way:
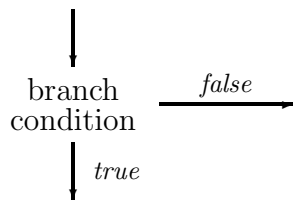
\begin{center}
\unitlength=.01in
\linethickness{1.0pt}
\begin{picture}(200,100)
\put(50,100){\vector(0,-1){30}}
\put(50,50){\makebox(0,0)[cc]{\shortstack{branch \\ condition}}}
\put(95,50){\vector(1,0){70}}
\put(130,60){\makebox(0,0)[cc]{\footnotesize \em false}}
\put(50,30){\vector(0,-1){30}}
\put(60,15){\makebox(0,0)[lc]{\footnotesize \em true}}
\end{picture}
\end{center}
Execution of the program enters from the top, and a logical
``branch condition'' is evaluated.  If the branch condition
is true, execution proceeds along the downward branch; if false,
along the rightward branch.  This is plainly reversible, as long
as the evaluation of the branch condition is done in a reversible
way; there is no ambiguity in the execution of the reversed
program.

However, a simple join
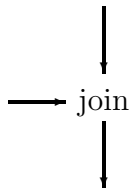
\begin{center}
\unitlength=.01in
\linethickness{1.0pt}
\begin{picture}(100,100)
\put(50,100){\vector(0,-1){35}}
\put(0,50){\vector(1,0){30}}
\put(50,40){\vector(0,-1){35}}
\put(50,50){\makebox(0,0)[cc]{join}}
\end{picture}
\end{center}
is not reversible, since in the reversed program it is not clear
which of the two paths to take.
The point is that a join in the
flowchart is a reversed branch, and thus must be governed by a
logical ``join condition'':
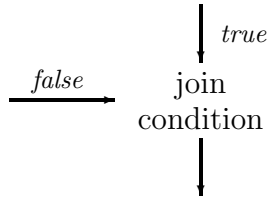
\begin{center}
\unitlength=.01in
\linethickness{1.0pt}
\begin{picture}(200,100)
\put(100,100){\vector(0,-1){30}}
\put(110,85){\makebox(0,0)[lc]{\footnotesize \em true}}
\put(100,50){\makebox(0,0)[cc]{\shortstack{join \\ condition}}}
\put(0,50){\vector(1,0){55}}
\put(25,60){\makebox(0,0)[cc]{\footnotesize \em false}}
\put(100,30){\vector(0,-1){30}}
\end{picture}
\end{center}
The program is designed so that the ``join condition'' is true
whenever the execution approaches from above, and false whenever
execution approaches from the right.

In our program, we will want to use branches and joins to
create ``loops'', like so:
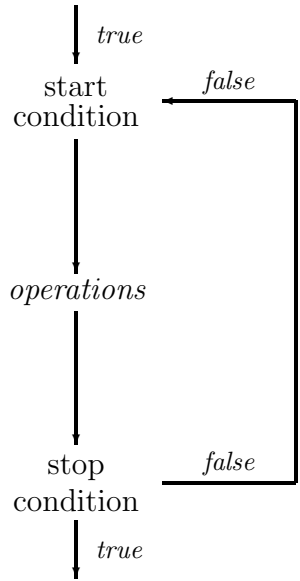
\begin{center}
\unitlength=.01in
\linethickness{1.0pt}
\begin{picture}(200,300)
\put(50,300){\vector(0,-1){30}}
\put(60,285){\makebox(0,0)[lc]{\footnotesize \em true}}
\put(50,250){\makebox(0,0)[cc]{\shortstack{start \\ condition}}}
\put(165,250){\vector(-1,0){70}}
\put(130,260){\makebox(0,0)[cc]{\footnotesize \em false}}
\put(50,230){\vector(0,-1){70}}
\put(50,150){\makebox(0,0)[cc]{\em operations}}
\put(50,140){\vector(0,-1){70}}
\put(50,50){\makebox(0,0)[cc]{\shortstack{stop \\ condition}}}
\put(50,30){\vector(0,-1){30}}
\put(60,15){\makebox(0,0)[lc]{\footnotesize \em true}}
\put(95,50){\line(1,0){70}}
\put(165,50){\line(0,1){200}}
\put(130,60){\makebox(0,0)[cc]{\footnotesize \em false}}
\end{picture}
\end{center}
The ``start condition'' is a logical condition that is only true at
the beginning of the execution of the loop and not thereafter; the
``stop condition'' is only true at the end of the execution of the
loop and not before.

We can also conveniently represent the
reversible loop structure in pseudocode form:
\begin{tabbing}
xxxxxxxx \= xxxx \= xxxx \= xxxx \= \kill
\> loop enter ({\em start condition}) \\
\>  \>  {\em operations} \\
\> loop exit ({\em stop condition})
\end{tabbing}
Both the beginning and the end of the loop are governed
by logical conditions.

The requirement that the computation be coherent may at first seem
difficult to achieve, since each branch point (or join point) in the
algorithm involves the evaluation of a condition---apparently a
measurement process.  However, these conditions can control the
execution of the program without any irreversible loss of coherence.

Let us suppose that the quantum system $Q$ is some portion of
our computer, and that we wish to branch our program based on
a condition about the state of $Q$.  The condition is represented
by a projection $\Pi$ acting on the Hilbert space $\hilbert$
describing $Q$.  Any initial state $\ket{\psi}$ of $Q$ can
be written as
\begin{equation}
	\ket{\psi} = \Pi \ket{\psi} + \Pi^\perp \ket{\psi}.
\label{start-branch-state-eqn}
\end{equation}
We could imagine evaluating the condition by making a measurement
of the observable represented by $\Pi$ and $\Pi^\perp$.  But
this would destroy the coherence in this superposition, so a
less destructive operation is required.

We join to $Q$ a single qubit (in another part of the quantum
computer), and consider the operator $U$ on joint system:
\begin{equation}
U = \left ( \ket{0}\bra{1} + \ket{1}\bra{0} \right )
\otimes \Pi + 1 \otimes \Pi^{\perp} .
\end{equation}
$U$ is easily verified to be a unitary operator, and thus
it could represent some coherent quantum evolution
of the joint system.  If the qubit is initially set to
the state $\ket{0}$ and then $U$ acts, we obtain,
\begin{equation}
U \ket{0} \otimes \ket{\psi} = \ket{1} \otimes \Pi \ket{\psi}
				+ \ket{0} \otimes \Pi^\perp \ket{\psi} .
\label{end-branch-state-eqn}
\end{equation}
This is an entangled state of the qubit and $Q$.  If we were
to make a measurement of the qubit in the standard basis, we
would be effectively measuring the observable $\Pi$ on $Q$.
That is, the qubit ``contains'' the value of the observable
$\Pi$.  However, the interaction is completely reversible,
and in this case may be undone by a further application of $U$
itself.

The qubit can be used as a switch to instruct the computer which
branch of the computation to follow.  Suppose we wish to specify
that, if the qubit is $\ket{0}$, the rest of the computer
performs a computation described by the unitary operator $V_0$,
whereas if the qubit is $\ket{1}$ then we wish to do the
computation $V_1$.  Then we instruct the entire computer
(including the switch qubit) to perform a coherent computation
described by the unitary operator
\begin{equation}
	V = \proj{0} \otimes V_0 + \proj{1} \otimes V_1 .
\end{equation}
If the overall state of the computer is a superposition
of the two switch states, both branches are followed
in different branches of the superposition.  The computer may
become increasingly entangled, but the coherence of its overall
state is preserved.

We have shown that any branching condition that can be represented by
a projection operator $\Pi$ can be used to control the execution of
the program without any necessary loss of coherence.  The cost is
entanglement among the parts of the computer.

A join point in the algorithm is simply a time-reversed branch point.
Just before the join, the computer is in
a state like  Equation~\ref{end-branch-state-eqn},
in which the qubit is entangled with the system $Q$.  The operator
$U^{-1} = U$ acts, and we return the qubit to the state $\ket{0}$
and the system $Q$ to a state like Equation~\ref{start-branch-state-eqn}.
We have ``disentangled'' $Q$ from the qubit, so the two branches of the
computation (controlled by the qubit) have merged.

Our second concern with coherent computation is the synchronization of
the computation on different components of the initial superposition.  This
can be maintained without much difficulty by introducing appropriate ``delay
loops'' into the program, so that its execution requires exactly the same
number of machine cycles for any input.

We will address our final concern, that the output tape should wind up
unentangled with the rest of the computer, by showing that the final
state of the rest of the computer (input registers and central processor)
is independent of the input state.

\subsection{Prefix-free codes are simply condensable}  \label{pf-is-simpcond-sec}

We are at last ready to give our algorithm for simply condensing the
codewords of a prefix-free quantum code.  First, we establish our
notation and describe the contents of our computer in slightly more
detail:
\begin{description}
\item[Registers]
Our computer contains $N$ registers, each consisting of $l_{max}$ qubits.
The $i$th register is denoted $R_i$ and the $k$th qubit of this register
is called $R_{i,k}$.  Initially, each register contains a \zef codeword
from a fixed prefix-free quantum code.
\item[Tape]
There is a ``tape'' $T$ containing at least $N l_{max}$ qubits, all of which
are initially in the state $\ket{0}$.  The $n$th qubit in the tape is called
$T_n$.
\item[Counter]
There is a counter variable $c$, which can take on integer values starting with
0 (or, of course, superpositions of these).  The initial state of $c$ is $\ket{0}$.
\item[Pointers]
There are several pointer variables, which like the counter variable take on
integer values and have an initial state $\ket{0}$.  These variables point to
locations in the computer's memory, but of course they are themselves quantum
variables and can take on entangled superpositions of values.  There is an
overall register pointer $r$ and, for each register, a qubit pointer $q_i$ (for
the $i$th register).  The tape also has a pointer variable $p$.
\end{description}

The first section of the program copies the contents of the registers
to the tapes, moving the pointers in the process.
\begin{tabbing}
xxxxxxxx \= xxxx \= xxxx \= xxxx \= \kill
\>       loop enter ($r = 0$)		\\
\>  \>      $r \leftarrow r + 1$		\\
\>  \>	    loop enter ($q_r = 0$)	\\
\>  \>  \>     $q_r \leftarrow q_r + 1$  \\
\>  \>  \>     $p \leftarrow p + 1$	\\
\>  \>  \>     $T_p \leftarrow T_p \oplus R_{r,q_r}$  \\
\>  \>      loop exit ($R_{r,1} \cdots R_{r,q_r}$ is a codeword of length $q_r$)  \\
\>       loop exit ($r = N$)
\end{tabbing}
(The notation $a \leftarrow a \oplus b$ indicates the ``controlled not''
operation on the qubits, with $a$ as the ``target'' and $b$ as the
``control'' qubit.)
Notice that the exit condition for the inner loop (which copies the
register qubits one by one onto the tape) is legitimate because the
code is prefix-free.  This means that the question of whether the
first $q_r$ qubits of the register form a codeword of length $q_r$
can be settled by measuring a projection-type observable on those
qubits.  (The computer does not make such a measurement, but instead
coherently controls its operation as described above.)

We also note that, since the procedure is just to copy the register
contents to the output tape, we are doing simple condensation.

At this stage, the various pointer variables are entangled with the codeword
length information; furthermore, the time at which the computer reaches this
stage of the computation is indeterminate.  We now resynchronize the
program via a delay loop that causes the computer to ``idle'' until a
fixed time $D$ (chosen large enough so that the first section of the program
has finished even for the longest possible input codewords).
\begin{tabbing}
xxxxxxxx \= xxxx \= xxxx \= xxxx \= \kill
\>        loop enter ($c = 0$)		\\
\>  \>       $c \leftarrow c + 1$	\\
\>	  loop exit ($\mbox{time} = D$)
\end{tabbing}

The second half of the program is the reverse of the first half,
except that the register is uncopied, rather than the tape.
\begin{tabbing}
xxxxxxxx \= xxxx \= xxxx \= xxxx \= \kill
\>        loop enter ($\mbox{time} = D+1$)		\\
\>  \>       $c \leftarrow c - 1$	\\
\>	  loop exit ($c = 0$) \\
\>       loop enter ($r = N$)		\\
\>  \>	    loop enter ($R_{r,1} \cdots R_{r,q_r}$ is a codeword of length $q_r$)  \\
\>  \>  \>     $R_{r,q_r} \leftarrow T_p \oplus R_{r,q_r}$  \\
\>  \>  \>     $p \leftarrow p - 1$	\\
\>  \>  \>     $q_r \leftarrow q_r - 1$ \\
\>  \>      loop exit ($q_r = 0$) 	\\
\>  \>      $r \leftarrow r - 1$	\\
\>       loop exit ($r = 0$)
\end{tabbing}
The program now ends, after exactly $2D$ machine steps.  All pointers and
counters have been returned to their initial zero values, and the
input qubit registers have been reset to $\ket{00 \cdots 0}$.
Only the qubit tape now contains any non-zero data, in the form of
a simply condensed string of $N$ codewords.  In short, the computer at
the end retains no codeword-length information at all.
Superpositions of codewords of different length will thus
remain coherent in the condensation process.  Since the algorithm
works for any given $N$, the prefix-free quantum code is
simply condensable.

We previously proved that every condensable code satisfies the quantum
Kraft-McMillan inequality, and then that every quantum code that
satisfies the Kraft-McMillan inequality can be unitarily remapped to
a prefix-free code.  We now learn that prefix-free quantum codes are
simply condensable.  Since unitary remapping might be part of a general
condensation process, we have established that a quantum code is
condensable if and only if it satisfies the quantum Kraft-McMillan
inequality.

\section{Quantum data compression}

\subsection{How many qubits?}

Classical variable-length codes are used for {\em data compression}---that is,
the representation of classical information in a compact way, using as few
resources (bits) as possible.  This is done by encoding more probable messages
in shorter codewords, so that the average codeword length is minimized.  In
this section we will discuss how---and in what sense---quantum
indeterminate-length codes may be used for quantum data compression.

Suppose Alice is sending classical information to Bob using the following
classical variable-length code:
\begin{center}
\begin{tabular}{cc}
	message		&	codeword	\\
	$C_1$		&	0	\\
	$C_2$		&	10	\\
	$C_3$		&	110	\\
	$C_4$		&	111
\end{tabular} .
\end{center}
If the message $C_1$ is sent, Bob receives a signal consisting of a single bit
(0); but if $C_4$ is sent, he receives three bits (111).  In each case,
Bob knows how many bits are being used to send the message.  If a long
string of messages is being sent, Bob at any stage knows how many complete
messages have been received.

Bob learns the length of each codeword because he actually learns which
codeword was sent.  The fact that Bob learns the identity of each codeword
is not a problem in the classical situation; indeed, it is the whole point
of classical communication!  This contrasts with quantum information transfer.
If Alice's signals, for example, are drawn from a non-orthogonal set of
states, Bob will not be able to determine reliably which signal was sent,
and any attempt to do so would damage the fidelity of the quantum information.

Suppose that Alice wishes to send quantum information to Bob using the
quantum analogue of the prefix-free code shown above.  In other words,
the length eigenstate \zef codewords are
\begin{center}
\begin{tabular}{lc}
	state		&	length	\\
	$\ket{000}$	&	1	\\
	$\ket{100}$	&	2	\\
	$\ket{110}$	&	3	\\
	$\ket{111}$	&	3
\end{tabular} .
\end{center}
Arbitrary superpositions of these codewords are also allowed codewords.
To maintain the coherence of these superpositions, therefore, Bob must
not obtain any information about the length of the codeword he receives.

A quantum system actually used for the transmission of information must
have at least two degrees of freedom.  The first is the ``data'' degree
of freedom, which may for instance be a qubit.  The second degree of freedom
is the ``location'' degree of freedom.  This is the physical degree of
freedom which determines whether or not Bob has access to the data degree
of freedom.  The faithful transmission of a qubit in a state $\ket{\psi}$
from Alice's location $a$ to Bob's location $b$ would be a process like
this:
\begin{equation}
	\ket{\psi,a} \rightarrow \ket{\psi,b} .
\end{equation}
Although we are phrasing our discussion in terms of the transmission of
quantum information from one spatial location to another, this analysis
would also apply to the storage and retrieval of information in a quantum
computer.  There the ``location'' degree of freedom might be the reading
of a clock; the information stored at time $a$
is to be retrieved at some later time $b$.

If we have several data qubits, each one will have a location degree of
freedom (which may, of course, be correlated with the others).
The number of qubits
transmitted from Alice to Bob will be the number of location degrees of
freedom that have evolved from $a$ to $b$.  For instance, suppose that three
data qubits are in a joint state $\ket{\psi^{123}}$, and that
Alice sends the first and third qubits to Bob.  The final state
would be $\ket{\psi^{123},bab}$, in which Bob has received two qubits.

How could Alice send an {\em indeterminate} number of qubits to Bob---in
particular, if Alice is representing her quantum information using the
prefix-free quantum code above, how can she arrange to send only the
first $l$ qubits of a \zef codeword of length $l$?  The transmission
of the length eigenstates is easy to describe:
\begin{eqnarray*}
\ket{000,aaa} & \rightarrow & \ket{000,baa} \\
\ket{100,aaa} & \rightarrow & \ket{100,bba} \\
\ket{110,aaa} & \rightarrow & \ket{110,bbb} \\
\ket{111,aaa} & \rightarrow & \ket{111,bbb}  .
\end{eqnarray*}
But imagine that Alice is sending a superposition of codewords of different
lengths.  If the above process is unitary, then at the end the data qubits
will be entangled with their location degrees of freedom.  The coherence of
the superposition would no longer be maintained within the data qubits.
In order to restore the coherence, Bob would have to interact with the
location degrees of freedom of the qubits with which he has indeterminate
access.  Except for a trivial case---in which Bob simply returns the
qubits from location $b$ back to $a$---he will not be able to do this.

If the transmission process is not unitary, things are even worse.
Our conclusion is that it is not possible to send quantum information
coherently using an indeterminate number of qubits.
If we are to use indeterminate-length quantum codes for quantum data
compression, we will have to do so in such a way that a {\em fixed}
number of qubits changes hands from Alice to Bob.

Perfect fidelity would demand that Alice send {\em all} of the qubits
to Bob---enough qubits so that even the longest component of each
codeword is transmitted in its entirety.
But this scheme would allow for no data compression at all.

Our previous discussion of condensability offers some hope.  The condensation
process took the ``information-bearing'' parts of $N$ \zef codewords (in
registers of length $l_{max}$) and unitarily shifted them as far
as possible toward the beginning of a tape of $N l_{max}$ qubits.
Although some branches of the overall superposition may extend to the
end of the tape, the ``typical'' branch may be much shorter (followed by
$\ket{0}$'s).  We therefore might be able to truncate the condensed string
of codewords after some number $L$ of qubits, where $L \ll N l_{max}$, and
still maintain an average fidelity approaching unity.

Let us consider a quantum information source that
produces an ensemble of signal states of some quantum system.
These signal states are unitarily encoded as
\zef codewords of some condensable quantum code.  For our purposes,
therefore, we can simply consider the ensemble of \zef codewords produced
by the quantum information source and the unitary encoding.  In this
ensemble, the codeword $\ket{a_{\subzef}}$ occurs with probability $p(a)$, and
the average encoded signal state is described by the density operator
\begin{equation}
	\rho = \sum_a p(a) \proj{a_{\subzef}} .
\end{equation}
Our source produces a sequence of independent, identically distributed
signals, which are encoded as \zef codewords in separate registers.  The
average state of $N$ of these registers is $\rho^{\otimes N}$.

The average length $\ave{l}$ of the codeword ensemble is
\begin{equation}
	\ave{l} = \tr \rho \lenobs = \sum_a p(a) \bra{a_{\subzef}} \lenobs \ket{a_{\subzef}} .
\end{equation}
The average length $\ave{l}$ is an ensemble average of quantum expectation
values for $\lenobs$, but no codeword $\ket{a_{\subzef}}$
need be a length eigenstate.

A condensed string of $N$ codewords is a \zef string of $N l_{max}$ qubits,
with length observable $\biglenobs$.  If $U$ is the unitary operator that
maps the $N$ separate \zef codewords to the condensed string, then
we can define the overall length observable for the condensed string to be
\begin{displaymath}
	\biglenobs = U \left ( \lenobs_1 + \lenobs_2 + \cdots + \lenobs_N \right ) U^{-1} .
\end{displaymath}
The condensed length $\biglenobs$ is just the sum of the individual length
observables of the separate, pre-condensed codewords.
This observable will have eigenvalues $L = l_1 + \cdots + l_N$
and an average value $\ave{L}$.  The codewords are independent, and so
\begin{equation}
	\ave{L} = N \ave{l} .
\end{equation}

Since the overall length of the condensed string is defined to be additive,
we can apply the ``law of large numbers'' to some measurement of $\biglenobs$:
For any $\epsilon, \delta > 0$, for large enough $N$ it is true that
\begin{equation}
	\prob{ | \biglenobs - N \ave{l} | > N \delta} < \epsilon .
	\label{LLN-eq}
\end{equation}
This means that, for large $N$, the probability is very small
that $\biglenobs$ will be found to be much less than (or much greater than)
$\ave{L}$.  Of course, we will not in general make such a measurement, but
Equation~\ref{LLN-eq} is still useful in restricting the typical
amplitude of codeword string components.

As we shall see, if the ensemble average length
of the \zef codewords is $\ave{l}$, then we can in the long run maintain
fidelity near to 1 by keeping just $\ave{l} + \delta$ qubits per signal,
where $\delta$ can be made as small as desired.  Conversely, in a simple
condensation process, we must keep at least $\ave{l}$ qubits per signal
to maintain high fidelity---if we keep only $\ave{l} - \delta$ per signal,
the average fidelity tends toward zero.  We will also find that the
ensemble average length of the \zef codewords is related to the von Neumann
entropy of the signal ensemble, making this approach an alternate route
to the noiseless quantum coding theorem.  Finally, we will show that the
relative entropy is a measure of the additional resources (qubits) required
to represent quantum information using a code that is not optimal.

\subsection{Enough qubits}

In this section we will make use of the fact that a condensed string
of $N$ \zef codewords is itself in \zef form---in other words, we can
view the condensed string as a \zef codeword in a much longer code.
The length observable for this super-codeword will be the sum of the
length observables for the $N$ original codewords.

Suppose we have a \zef codeword $\ket{\phi}$ in a register of $n$
qubits, and suppose $\ell \leq n$.  Define $\eta$ such that
a measurement of the length observable $\Lambda$ on the codeword
yields a result larger than $\ell$ with probability
\begin{equation}
	\prob{\Lambda > \ell} = \eta .
\end{equation}
In general $\ket{\phi}$ will include components of various lengths.
Let $\Pi_{\ell}$ be the projection
\begin{equation}
	\Pi_{\ell} = 1^{1 \cdots \ell} \otimes \proj{0^{\ell+1 \cdots n}} .
\end{equation}
That is, $\Pi_{\ell}$ projects onto the subspace of register states that
are $\ket{0}$ in the last $n-\ell$ qubits.  We can write our \zef codeword
$\ket{\phi}$ as
\begin{equation}
	\ket{\phi} = \alpha \ket{\phi\zshortish} + \beta \ket{\phi\zlong}
\end{equation}
where $\alpha, \beta \geq 0$, and $\ket{\phi\zshortish}$ and $\ket{\phi\zlong}$
are normalized states such that
\begin{eqnarray*}
	\Pi_{\ell} \ket{\phi \zshortish} & = & \ket{\phi \zshortish} \\
	\Pi_{\ell} \ket{\phi \zlong} & = & 0 .
\end{eqnarray*}

Since all $\Lambda$-eigenstate \zef codewords
with length no larger than $\ell$
have $\ket{0}$ in the last $n-\ell$ qubits,
\begin{equation}
	1 - \eta = \prob{\Lambda \leq \ell} \leq \alpha^2 .
\end{equation}
Equality need not hold, however, since some length eigenstate
codewords with $\Lambda > \ell$ may nevertheless have $\ket{0}$ in
the last $n - \ell$ qubits.  (This is analogous to the classical
situation, in which it is perfectly possible to have one or more 0's
at the end of a codeword in a variable-length code.)

We now imagine that we truncate the register by discarding the
last $n - \ell$ qubits.  Only $\ell$ qubits are stored or transmitted.
At the receiver's end of the process, $n - \ell$ qubits in the
standard state $\ket{0}$ are appended, yielding a mixed
final state $\sigma$ for the register.
With what fidelity $F = \bra{\phi} \sigma \ket{\phi}$
has the original codeword state been maintained by this process?

Direct calculation shows that the mixed state $\sigma$ is
\begin{equation}
	\sigma = \alpha^2 \proj{\phi\zshortish} + \beta^2 w\zlong
\end{equation}
where $w\zlong$ is the state obtained by truncating $\ket{\phi\zlong}$
and appending $n - \ell$ qubits in the state $\ket{0}$.  Thus
\begin{eqnarray}
	F & = & \bra{\phi} \sigma \ket{\phi} \nonumber \\
	  & = & \alpha^2 \left | \amp{\phi}{\phi\zshortish} \right |^2 +
			\beta^2 \bra{\phi} w\zlong \ket{\phi}  \nonumber \\
	  & \geq & \alpha^4 .
\end{eqnarray}
Therefore,
\begin{equation}
	F \geq \alpha^4 \geq (1-\eta)^2 \geq 1 - 2 \eta .
\end{equation}
If the codeword length $\Lambda$ would be found to be no more than $\ell$
with probability $1 - \eta$, then we can keep only $\ell$ qubits and
recover the original state with fidelity $F \geq 1 - 2 \eta$.

We can now apply this result and the law of large numbers (Equation~\ref{LLN-eq})
to a condensed string of codewords.  If $\epsilon, \delta > 0$ and $N$ is
sufficiently large, and if we take $\ell = N(\ave{l} + \delta)$,
then the ensemble average probability that the codeword string is longer than $\ell$
can be made smaller than $\epsilon / 2$.  We can therefore truncate the string
after only $N(\ave{l} + \delta)$ qubits and later recover the original string
with an average fidelity
\begin{equation}
	\ave{F} > 1 - \epsilon.
\end{equation}
Therefore, if we keep more than $\ave{l}$ qubits per input message,
in the long run we will be able to retrieve the quantum information
with average fidelity approaching unity.  The average length $\ave{l}$
tells us how many qubits are sufficient for high fidelity.

\subsection{Too few qubits}

We now turn to the question of how many qubits are necessary to achieve
high fidelity after the condensed string is truncated.  For this
discussion we will restrict our attention to {\em simple condensation},
rather than a general condensation process.  Since any condensable
code can be replaced by a simply condensable code with the same
length characteristics, this restriction is not too severe.

The reason for making this restriction is pragmatic.
Suppose we have $N$ registers containing codewords from a
condensable code, with an average length of $\ave{l}$.
A general condensation procedure might consist of two stages.
In the first, the codewords in the $N$ separate registers are
unitarily remapped to codewords from a more efficient code, that
is, one with shorter average length $\ave{l'} < \ave{l}$.
In the second stage, this more efficient code is condensed.
We have established that only about $N \ave{l'}$ qubits will be
sufficient to maintain high fidelity.  In other words, the
original average length $\ave{l}$ may tell us nothing about
the number of qubits necessary for high fidelity.

Of course, we might not choose to condense the codewords in this
way, or a more efficient code might not exist.  Our strategy
will be to separate the question of the efficiency of a code from
the question of how many qubits are necessary.  First we will
consider the simple condensation of codes that may be inefficient,
and then (in the next section) we will discuss limits on the efficiency
of codes.  In this section, therefore, we describe limits
imposed by the structure of our particular (possibly sub-optimal)
code, and in the next we will indicate how optimal or near-optimal
codes may be chosen.

Begin with $N$ \zef codewords of a simply condensable code.  The simply condensed
string formed from the $N$ codewords can be built out of two pieces:
\begin{enumerate}
	\item  the simply condensed qubit string obtained from the first $N - k$
		codewords, and
	\item  the simply condensed qubit string obtained from the last $k$ codewords.
\end{enumerate}
These two pieces are both \zef and are simply condensed together
to form the complete string.  Thus, we will base our discussion on the simple
condensation of just two \zef codewords.

The first \zef codeword $\ket{\psi}$ lies in a register of $m$ qubits, and
the second codeword $\ket{\chi}$ lies in a register of $n$ qubits.  The
simply condensed pair (denoted rather symbolically by $\ket{\psi \chi}$)
is a state of a string of $m + n$ qubits.  We also consider a state called
$\ket{\psi 0}$, which is the first \zef codeword followed by $n$ additional
qubits in the state $\ket{0}$.

Let $\ell \leq m + n$.  The first \zef codeword can be written
\begin{equation}
	\ket{\psi} = \alpha \ket{\psi\oshort} + \beta \ket{\psi\olongish}
\end{equation}
where $\alpha, \beta \geq 0$ and $\ket{\psi\oshort}$ (or $\ket{\psi\olongish}$)
is a normalized superposition of length eigenstates that are
shorter than (or at least as long as) $\ell$.
If we now simply condense this codeword with the codeword $\ket{\chi}$,
we obtain
\begin{equation}
	\ket{\psi \chi} = \alpha \ket{\psi\oshort \chi} + \beta \ket{\psi\olongish \chi},
\end{equation}
with $\ket{\psi\oshort \chi}$ and $\ket{\psi\olongish \chi}$ being the simply condensed
strings obtained from $\ket{\chi}$ and the two components of $\ket{\psi}$.  In
a similar way,
\begin{equation}
	\ket{\psi 0} = \alpha \ket{\psi\oshort 0} + \beta \ket{\psi\olongish 0}.
\end{equation}

Now we imagine truncating the string of $m+n$ qubits, keeping only the first
$\ell$ of them to be stored or transmitted.  (We can denote this process
by $\truncate_{\ell}$.)  At the receiver's end, we do some
unspecified quantum operation $\superop$ that results in a final state of
$m+n$ qubits.  We know nothing about $\superop$ in general except that it is
a trace-preserving, completely positive linear map on density operators.
The overall process, applied to the two initial states $\ket{\psi \chi}$
and $\ket{\psi 0}$, yield
\begin{eqnarray*}
  \ket{\psi \chi}  &  \stackrel{\truncate_{\ell}}{\longrightarrow} \omega
			\stackrel{\superop}{\longrightarrow}  &  \superop(\omega) \\
 \ket{\psi 0}  &  \stackrel{\truncate_{\ell}}{\longrightarrow} \sigma
			\stackrel{\superop}{\longrightarrow}  &  \superop(\sigma) .
\end{eqnarray*}
At the end of this process, we are interested in the overall fidelity of the
truncation-{\em cum}-recovery process:
\begin{equation}
	F = \bra{\psi \chi} \superop(\omega) \ket{\psi \chi} .
\end{equation}
We will show that, under suitable conditions,
this fidelity must be small.

For general density operators, the fidelity is defined to be
\begin{equation}
	F(\rho_1,\rho_2) = \max | \amp{1}{2} |^2 ,
\end{equation}
where the maximum is taken over all purifications $\ket{1}$ of $\rho_1$ and
$\ket{2}$ of $\rho_2$.  (Equivalently, we can fix one of the purifications
$\ket{1}$ and maximize over the other purification $\ket{2}$.)  The fidelity
has the property that it is never decreased by any quantum operation, so that
\begin{equation}
	F \left ( \superop(\rho_1) , \superop(\rho_2) \right )
		\geq F(\rho_1,\rho_2)
\end{equation}
for any trace-preserving, completely positive linear map $\superop$.

A useful result (shown in \cite{fidelity-lemma})
relates the fidelities among three states $\rho_1$,
$\rho_2$ and $\rho_3$.  Let $F_{12} = F(\rho_1,\rho_2)$, etc.  Then
\begin{equation}
	\sqrt{F_{13}} \leq \sqrt{F_{23}} + \sqrt{2 ( 1 - \sqrt{F_{12}})}  .
\end{equation}
This implies that, if $F_{12}$ is nearly equal to one and $F_{23}$ is
close to zero, $F_{13}$ is also close to zero.  Recalling that $0 \leq F \leq 1$
for all fidelities, we note that $1-\sqrt{F} \leq 1 - F$, and thus
\begin{eqnarray}
	F_{13} & \leq &	F_{23} + 2(1 - F_{12}) + 2 \sqrt{2 F_{23}(1 - F_{12}) }
				\nonumber \\
		& \leq & F_{23} + 2(1 - F_{12}) + 2 \sqrt{2 (1 - F_{12}) }
				\nonumber \\
		& \leq & F_{23} + 2\sqrt{1 - F_{12}} + 2 \sqrt{2} \sqrt{1 - F_{12} }
				\nonumber \\
	F_{13}  & \leq & F_{23} + 5 \sqrt{1-F_{12}} .
		\label{eq-fidineq}
\end{eqnarray}
Since this inequality is linear in both $F_{13}$ and $F_{23}$, it will be
convenient for situations in which we wish to average over an ensemble of
$\rho_3$ states.

We apply Equation~\ref{eq-fidineq} to our situation as follows.  The state
$\rho_1 = \proj{\psi \chi}$, the original simply condensed string, and the
state $\rho_3 = \superop(\omega)$, the final state of the simply condensed
string after the truncation $\truncate_{\ell}$ and
the recovery operation $\superop$.
Playing the role of $\rho_2$ is the state $\superop(\sigma)$, the final
state obtained by using $\ket{\psi 0}$ as our input.  Since the quantum
operation $\superop$ can never decrease the fidelity between states,
$F(\superop(\omega),\superop(\sigma)) \geq F(\omega,\sigma)$.  Therefore,
\begin{eqnarray}
F & = & \bra{\psi \chi} \superop(\omega) \ket{\psi \chi} \nonumber \\
  & \leq &
	\bra{\psi \chi} \superop(\sigma) \ket{\psi \chi} +
	5 \sqrt{1 - F(\omega,\sigma)} .
\end{eqnarray}

The initial states $\ket{\psi \chi}$ and $\ket{\psi 0}$ are purifications
of $\omega$ and $\sigma$, respectively.  The fidelity $F(\omega,\sigma)$ is
thus
\begin{equation}
	F(\omega,\sigma) = \max_{\ket{\phi_{\sigma}}}  \left |
		\amp{\psi \chi}{\phi_{\sigma}} \right |^2
\end{equation}
where the maximum is taken over all purifications $\ket{\phi_{\sigma}}$ of
$\sigma$.  Now, all of the purifications of $\sigma$ are related to one another
by unitary operators that act only on the adjoined system, so that
\begin{equation}
	F(\omega,\sigma) =  \max_{U}
			\left | \bra{\psi \chi} \left ( 1^{1 \cdots \ell} \otimes
			U^{\ell + 1 \cdots m+n} \right ) \ket{\psi 0} \right |^2 .
\end{equation}
with the maximum taken over all unitary operators acting on the last $m+n-\ell$
qubits.

We write $\ket{\psi \chi} = \alpha \ket{\psi\oshort \chi} + \beta \ket{\psi\olongish \chi}$
and $\ket{\psi 0} = \alpha \ket{\psi\oshort 0} + \beta \ket{\psi\olongish 0}$, as before,
and note that, since $\ket{\psi\olongish}$ only contains components of $\ket{\psi}$
that are at least as long as $\ell$,
\begin{equation}
	\tr_{\ell+1 \cdots m+n} \proj{\psi\olongish \chi}
		=  \tr_{\ell+1 \cdots m+n} \proj{\psi\olongish 0} .
\end{equation}
In this component,
the second codeword, whose ``starting address'' in the simply condensed
string is entangled with the length of the first codeword,
lies entirely in the discarded tail of the qubit string.
Therefore, there exists a unitary $V^{\ell + 1 \cdots m+n}$ such that
\begin{equation}
	\ket{\psi\olongish \chi} = \left ( 1^{1 \cdots \ell} \otimes
		V^{\ell + 1 \cdots m+n} \right )\ket{\psi\olongish 0} .
\end{equation}
Clearly,
\begin{equation}
	F(\omega,\sigma)
	  \geq  \left | \bra{\psi \chi} \left ( 1^{1 \cdots \ell} \otimes
			V^{\ell + 1 \cdots m+n} \right ) \ket{\psi 0} \right |^2 .
\end{equation}

\begin{eqnarray*}
  \bra{\psi \chi} \left ( 1^{1 \cdots \ell} \otimes
			V^{\ell + 1 \cdots m+n} \right ) \ket{\psi 0} & = &
	\alpha^2 \bra{\psi\oshort \chi} \left ( 1^{1 \cdots \ell} \otimes
			V^{\ell + 1 \cdots m+n} \right ) \ket{\psi\oshort 0} \\
  &  & + \alpha \beta \bra{\psi\oshort \chi} \left ( 1^{1 \cdots \ell} \otimes
			V^{\ell + 1 \cdots m+n} \right ) \ket{\psi\olongish 0} \\
  &  & + \beta \alpha \bra{\psi\olongish \chi} \left ( 1^{1 \cdots \ell} \otimes
			V^{\ell + 1 \cdots m+n} \right ) \ket{\psi\oshort 0} \\
  &  &  + \beta^2 \\
\left | \bra{\psi \chi} \left ( 1^{1 \cdots \ell} \otimes
	V^{\ell + 1 \cdots m+n} \right ) \ket{\psi 0} \right |
  & \geq &  1 - 2 \alpha - 2 \alpha^2  \geq 1 - 4 \alpha .
\end{eqnarray*}
Therefore
\begin{eqnarray}
	F(\omega,\sigma) & \geq & ( 1 - 4 \alpha )^2 \nonumber \\
			& \geq &  1 - 8 \alpha .
\end{eqnarray}
Our overall fidelity must satisfy
\begin{eqnarray}
	F & \leq & \bra{\psi \chi} \superop(\sigma) \ket{\psi \chi} +
		5 \sqrt{8 \alpha} \nonumber \\
	  & \leq & \bra{\psi \chi} \superop(\sigma) \ket{\psi \chi} +
		15 \sqrt{\alpha} .
\end{eqnarray}

Neither the operator $\superop(\sigma)$ nor the parameter $\alpha$
depends on the second codeword $\ket{\chi}$.  We now imagine that
the second codeword is drawn from an ensemble---that is, that the
codeword $\ket{\chi}$ occurs with probability $P(\chi)$, so that
the ensemble has an average density operator
\begin{equation}
	W = \sum_{\chi} P(\chi) \, \proj{\chi} .
\end{equation}
The average fidelity after truncation $\truncate_{\ell}$ and
recovery $\superop$ will therefore be
\begin{equation}
	\bar{F} \leq \tr W \superop(\sigma) + 15 \sqrt{\alpha}.
\end{equation}
Since $\superop(\sigma)$ is a positive operator of unit trace, we
obtain
\begin{equation}
	\bar{F} \leq \opnorm{W} + 15 \sqrt{\alpha} , \label{eq-barfest}
\end{equation}
where $\opnorm{W}$ is the operator norm of $W$, which (since $W$ is
positive) is just the largest eigenvalue of $W$.

After all of this, we are in a position to apply the law of large
numbers (Equation~\ref{LLN-eq}) again.  We will be choosing
two large integers, $N$ and $k$.  Our first codeword $\ket{\psi}$
in the preceding analysis will be a simply condensed string of
$N-k$ codewords, and the second codeword $\ket{\chi}$ will be a simply condensed
string of the remaining $k$ codewords.  We assume that the ensemble
of single-register codewords has an average state $\rho$ with more than
one non-zero eigenvalue---in other words, the ensemble involves more
than one codeword state.

Let $\epsilon, \delta > 0$.  If $\lambda < 1$ is the largest eigenvalue
of $\rho$, then the largest eigenvalue of $\rho^{\otimes k}$ is $\lambda^{k}$.
Choose $k$ so that $\lambda^k < \epsilon/2$.
Since the last $k$ codewords are unitarily condensed into a string
with average state $W$, $\opnorm{W} = \opnorm{\rho^{\otimes k}} < \epsilon/2$.

Now we consider the simply condensed string of the first $N-k$ codewords,
which we have denoted by $\ket{\psi}$.  The length observable for this
string is $\lenobs_{N-k}$.  Given a value of $N$, we define
$\ell = N(\ave{l} - \delta)$.
We will restrict our attention to values of $N$ large enough so that
\begin{equation}
	\ell \leq (N-k) \left( \ave{l} - \frac{\delta}{2} \right ) .
\end{equation}
Applying the law of large numbers, we can now specify $N$ large enough so that
$\prob{\lenobs_{N-k} < \ell} = \alpha^2$ is as small as we like.  In particular, we
can guarantee that $15 \sqrt{\alpha} < \epsilon / 2$.  Thus,
\begin{equation}
	\bar{F} \leq \opnorm{W} + 15 \sqrt{\alpha} < \epsilon .
\end{equation}

Therefore, if we keep fewer than $\ave{l}$ qubits per input message
and use simple condensation, in the long run the fidelity of the
retrieved quantum information must approach zero.  The average length
$\ave{l}$ tells us how many qubits are necessary for high fidelity
using simple condensation.

\subsection{Entropy and average length}

The preceeding results provide an interpretation for the average length
$\ave{l}$ of an indeterminate-length quantum code:  $\ave{l}$ is just
a measure of the resources (qubits) that are both necessary and sufficient
to maintain high fidelity of the quantum information, in the situations
described above.  We now inquire how short $\ave{l}$ can be for a given
quantum information source.  In other words, we will now explore how
efficient an indeterminate-length quantum code may be.

Recall the quantum Kraft-McMillan inequality (Equation~\ref{kraft}).
Any condensible quantum code must have a length observable $\lenobs$
on \zef codewords that satisfies
\begin{displaymath}
	\tr 2^{-\lenobs} = K \leq 1 .
\end{displaymath}
where the trace is restricted to the \zef subspace.
We can construct a density operator $\omega$
on the \zef subspace by letting
\begin{equation}
	\omega = \frac{1}{K} \, 2^{-\lenobs} .
\end{equation}
The operator $\omega$, although a positive operator of
unit trace, is generally not the same as the ensemble
average density operator $\rho$ of the codewords
produced by the information source.

The average codeword length $\ave{l}$ is
\begin{eqnarray*}
	\ave{l} & = & \tr \rho \lenobs \\
		& = & - \tr \rho \log \left ( 2^{-\lenobs} \right ) \\
		& = & - \tr \rho \log \omega - \log K .
\end{eqnarray*}
Therefore
\begin{equation}
	\ave{l}  =  S(\rho) + \relent{\rho}{\omega} - \log K ,
		\label{barleq}
\end{equation}
where $S(\rho)$ is the von Neumann entropy of the density operator $\rho$
\begin{equation}
	S(\rho) = - \tr \rho \log \rho
\end{equation}
and $\relent{\rho}{\omega}$ is the quantum relative entropy
\begin{equation}
	\relent{\rho}{\omega} = \tr \rho \log \rho - \tr \rho \log \omega .
\end{equation}
(We use base-2 logarithms.)  The relative entropy has a number of useful
properties.  For example, it is positive-definite, so that
$\relent{\rho}{\omega} > 0$ if and only if $\rho \neq \omega$.

Since $\log K \leq 0$,
\begin{equation}
	\ave{l} \geq S(\rho) .  \label{vonNeumannbound}
\end{equation}
The average codeword length
must always be at least as great as the von Neumann
entropy of the signal ensemble from the information source.

We can approach this bound by a suitable code.
The eigenvalues $\lambda_k$ of
$\rho$ form a probability distribution $\vec{\lambda}$,
and the von Neumann entropy is simply the Shannon entropy
of the eigenvalues:
\begin{equation}
	S(\rho) = H(\vec{\lambda}) = - \sum_k \lambda_k \log \lambda_k .
\end{equation}
The probability distribution $\vec{\lambda}$ can be used to
define a Shannon-Fano code, which is a classical prefix-free binary code
whose codewords have integer lengths $l_k = \lceil \log \lambda_k \rceil$,
so that
\begin{equation}
	l_k <  \log \lambda_k + 1 .
\end{equation}
This means that the average length of the Huffman codewords satisfies
\begin{equation}
	\ave{l} = \sum_k \lambda_k l_k < H(\vec{\lambda}) + 1 .
	\label{shannon-fano-bound}
\end{equation}
The classical Shannon-Fano code can be used to define a corresponding
prefix-free indeterminate-length quantum code, according to the
procedure in Equation~\ref{ccode2qcode}.  (Such a code was also
described by Chuang and Modha in \cite{chuang}.)
Eigenstates of $\rho$
are length eigenstate \zef codewords, and the average codeword
length satisfies
\begin{equation}
	\ave{l} < S(\rho) + 1 . \label{q-shannon-fano-bound}
\end{equation}
Asymptotically, this code will achieve high fidelity using about
$S(\rho) + 1$ qubits per signal.

An alternate scheme is based on Huffman codes, which are classical
prefix free codes that actually minimize average codeword length
$\ave{l}$.  Equations~\ref{shannon-fano-bound} and \ref{q-shannon-fano-bound}
are also satisfied for Huffman codes and their quantum versions.

We can do even better if we create our \zef codewords from {\em blocks}
of outputs of the quantum information source.  This amounts to considering
a new source that produces blocks of $n$ elementary signals,
with an ensemble average block state $\rho^{\otimes n}$ having
an entropy of $n S(\rho)$.  A quantum Shannon-Fano or Huffman code
designed for this block source would have an average length of no
more than $n S(\rho) + 1$, so that we will use only $S(\rho) + \frac{1}{n}$
qubits per elementary signal.  Thus, by coding long blocks of signals,
we can achieve $\bar{F} \rightarrow 1$
with about $S(\rho)$ qubits per elementary signal.

It can be seen that the theory of indeterminate-length quantum codes
provides an alternate route to the quantum noiseless coding theorem
\cite{qcompress}.  The von Neumann entropy $S(\rho)$
measures the physical resources necessary to represent
quantum information faithfully.

We now ask:  under what circumstances can we achieve the entropic
bound to the codeword length exactly, without resorting to
block coding?  In other words, for what codes and codeword
ensembles can we have
\begin{equation}
	\ave{l} = S(\rho) ?
\end{equation}
A code for which this equality holds may be called ``length optimizing''.
The answer can be seen from Equation~\ref{barleq}:
\begin{displaymath}
	\ave{l}  =  S(\rho) + \relent{\rho}{\omega} - \log K .
\end{displaymath}
Both $\relent{\rho}{\omega}$ and $- \log K$ are non-negative, so
they must both equal zero for a length optimizing code.  In other
words,
\begin{equation}
	K = \tr 2^{-\lenobs} = 1
\end{equation}
and
\begin{equation}
	\rho = \omega = 2^{-\lenobs} .
\end{equation}
A length optimizing code must saturate the quantum Kraft inequality
(Equation~\ref{kraft}), and the codeword ensemble must equal the
density operator $\omega$ constructed from the length observable $\lenobs$.
Two consequences follow:
\begin{itemize}
	\item  Whenever the signal ensemble $\rho$ has only eigenvalues
		of the form $2^{-m}$ for integer values of $m$, we can
		find a condensable quantum code (with length eigenvalues
		$m$) that is length optimizing.  If $\rho$ has eigenvalues
		that are not of this form, then no length optimizing
		code exists.
	\item  Some quantum codes saturate the quantum Kraft
		inequality---for example, those based on classical
		Huffman codes.  These codes will be length optimizing
		for a codeword ensemble with density operator
		\begin{equation}
			\rho = 2^{-\lenobs} .
		\end{equation}
		That is, every quantum code that saturates the quantum Kraft
		inequality is length optimizing for some codeword ensemble.
		If a quantum code does not saturate the quantum Kraft
		inequality, it is not length optimizing for any codeword
		ensemble.
\end{itemize}

Suppose we have a code that is length optimizing for some density operator
$\omega$; but instead, we use the code for an ensemble of codewords
described by the density operator $\rho$.  Then the average codeword
length will be
\begin{equation}
	\ave{l} = S(\rho) + \relent{\rho}{\omega} .
\end{equation}
We know that, using block coding, we can asymptotically use as few as
$S(\rho)$ qubits to faithfully represent the quantum information produced
by the source of $\rho$.  We also know that $\ave{l}$ is the minimum
number of qubits we need to retain per codeword to achieve high fidelity
in a simply condensed string of many codewords.  Thus, the relative
entropy $\relent{\rho}{\omega}$ tells us what additional resources
(in qubits) are necessary to faithfully represent the quantum information
from the $\rho$-source, if we use a code that is length optimizing for
a different source (the ``$\omega$-source'').

\subsection{Remarks}  \label{discuss-sec}

In the quantum Huffman code of Braunstein et al.,
codeword length information and the codewords themselves are stored
separately, in entangled strings of qubits.  This means that
the average number of qubits used to store the quantum
information from a given source is increased by an amount
logarithmic in the codeword length \cite{qhuffman}.
However, as we have seen, this separate accounting for
codeword length information is unnecessary.  The
codewords of a quantum indeterminate-length code {\em carry
their own length information}.

This requirement is the basis for Equation~\ref{kraft}, the quantum
Kraft-McMillan inequality.  We have shown that Equation~\ref{kraft} is
a necessary and sufficient condition for condensability, and further,
that any code satisfying Equation~\ref{kraft} can be unitarily mapped
to a prefix-free quantum code with the same length characteristics.
Prefix-free codes are themselves simply condensable, and obey the
quantum Kraft-McMillan inequality.

Classical prefix-free codes are also called ``instantaneous codes'',
since the receiver of a string of codewords can identify an individual
codeword from the string immediately, before the remainder of the
string is received \cite{cover}.
But this terminology is inapplicable to the quantum case.
Suppose we have a simply condensed string
of codewords from a prefix-free quantum code.  The first codeword
is generally not a length eigenstate, and the length of this
codeword is entangled with the locations in the qubit string
of all subsequent codewords.  The phase relationship
between the different-length components of the first codeword
is a global property of the state of the entire string.
Therefore, in order to coherently recover even the first
codeword, we will need the entire string (or a sufficiently
long initial segment to achieve high overall fidelity).
Even prefix-free quantum codes are not ``instantaneous'';
the entire transmission must be completed before any part
of it can be ``read''.

The classical Kraft-McMillan inequality (Equation~\ref{classkraft})
arises whenever a set of
binary strings satisfies the prefix-free condition.  For example, it
governs the set of lengths of distinct programs for
a classical Turing machine.
The Kraft-McMillan inequality therefore plays a central role in
{\em algorithmic} information theory, in which the information
content of a binary string $s$ is defined to be length of the
shortest halting program that produces $s$ as its output
\cite{cover, vitanyi}.
We may hope that the quantum version of the Kraft-McMillan
inequality will serve as a starting point for the development
of a quantum algorithmic information theory.

We are happy to acknowledge our indebtedness to many colleagues
with whom we have discussed this work, including C. M. Caves,
S. Braunstein, C. A. Fuchs, W. K. Wootters, T. M. Cover, and
I. L. Chuang.  One of us (BS) is grateful for the support of
a Rosenbaum Fellowship at the Isaac Newton Institute for
Mathematical Sciences in the summer of 1999.


\begin{thebibliography}{99}

\bibitem{chbinphystoday}  C. H. Bennett, {\em Physics Today} {\bf 48}, 24, (1995).
C. H. Bennett and D. P. DiVincenzo, {\em Nature} {\bf 404}, 247 (2000).
%
\bibitem{mikeandike}M. A. Nielsen and I. L. Chuang, {\em Quantum Computation
and Quantum Information} (Cambridge University Press, Cambridge, 2000).
%
\bibitem{cover}  T. M. Cover and J. A. Thomas, {\em Elements of Information Theory}
(John Wiley and Sons, New York, 1991).
%
\bibitem{beninsantafe} B. Schumacher, presentation at Sante Fe Institute workshop on
Complexity, Entropy and the Physics of Information (1994).
%
\bibitem{qhuffman}  S. L. Braunstein, C. A. Fuchs, D. Gottesman, and H.-K. Lo,
"A quantum analog of Huffman coding," in {\em Proceedings of the 1998 IEEE
International Symposium on Information, MIT, Cambridge, MA, USA, August 16-21,} page 353,
1998. http://xxx.lanl.gov/abs/quant-ph/9805080.
%
\bibitem{chuang}  I. L. Chuang and D. S. Modha, {\em IEEE Trans. Inf. Theory}
{\bf 46}(3), 1104 (2000).
%
\bibitem{bostrom}  K. L Bostr\"{o}m, http://xxx.lanl.gov/abs/quant-ph/0009052.
K. L Bostr\"{o}m, http://xxx.lanl.gov/abs/quant-ph/0009073.
%
\bibitem{qcompress}  B. Schumacher, {\em Physical Review A} {\bf 51}, 2738 (1995).
R. Jozsa and B. Schumacher, {\em J. Mod. Opt.} {\bf 41} 2343 (1994).
%
\bibitem{qerror} P. W. Shor, {\em Physical Review A}, {\bf 52}, 2493 (1995); A.R.
Calderbank and P. W. Shor {\em Physical Review A}, {\bf 54}, 1098 (1996); A. Steane
{\em Physical Review Letters}, {\bf 77}, 793 (1996); R. Laflamme, C. Miquel, J. P. Paz,
and W. H. Zurek, {\em ibid.} {\bf 77}, 198 (1996).
%
\bibitem{qcapacity}  C. H. Bennett, D. P. DiVincenzo, J. A. Smolin and
W. K. Wootters, {\em Phys Rev. A} {\bf 54} 3824 (1996).  H. Barnum,
M. A. Nielsen and B. Schumacher {\em Phys. Rev. A} {\bf 57}, 4153 (1998).
%
\bibitem{bennett}  C. H. Bennett, {\em IBM J. Res. Dev.} {\bf 17}, 525 (1973).
%
\bibitem{toffoli}  T. Toffoli, in {\em Automata, Languages, and Programming},
edited by W. de Bakker and J. van Leeuwen (Springer, New York, 1980).
%
\bibitem{fidelity-lemma} H. Barnum, C. A. Fuchs, R. Jozsa and B. Schumacher,
{\em Physical Review A} {\bf 54} 4707 (1996).
%
\bibitem{vitanyi}  Ming Li and P. Vitanyi, {\em An Introduction to Kolmogorov
Complexity and Its Applications (Second Edition)} (Springer-Verlag, Berlin, 1997).
%
\end{thebibliography}
\end{document}